\documentclass[prc,twocolumn,showpacs,preprintnumbers,amsmath,amssymb]{revtex4-1}
\usepackage[dvips]{graphicx}
\usepackage{dcolumn}
\usepackage{bm}

\begin{document}
\title{Effects of ground-state correlations on dipole and quadrupole excitations of $^{40}$Ca and $^{48}$Ca}
\author{Mitsuru Tohyama}
\affiliation{Kyorin University School of Medicine, Mitaka, Tokyo
  181-8611, Japan     }
\begin{abstract}
The effects of ground-state correlations on the dipole and quadrupole excitations are studied for $^{40}$Ca and $^{48}$Ca using the extended
random phase approximation (ERPA) derived from the time-dependent density-matrix theory. 
Large effects of the ground-state correlations are found in the fragmentation of the giant quadrupole resonance in $^{40}$Ca and in the low-lying 
dipole strength in $^{48}$Ca. It is discussed that the former is due to a mixing of different configurations in the ground state and the 
latter is from the partial occupation of the neutron single-particle states. 
The dipole and quadrupole strength distributions below 10 MeV calculated in ERPA are in qualitatively agreement with experiment.
\end{abstract}
\maketitle

\section{Introduction}
The random phase approximation (RPA) based on the Hartree-Fock (HF) ground state has extensively been used to study nuclear collective
excitations. It is generally considered that the HF + RPA approach is the most appropriate for doubly-closed shell nuclei such as $^{16}$O and $^{40}$Ca
for which the HF theory would give a good description of the ground states. 
However, it has also long been known\cite{zuker,agassi,kramer} that the ground states of $^{16}$O and $^{40}$Ca
are highly correlated.
Recent theoretical studies for $^{16}$O also show that the ground state of $^{16}$O is highly correlated and that the occupation probabilities of the single-particle
states near the Fermi level deviate more than 10 \% 
from the HF values \cite{Toh07,utsuno}. This indicates that the HF + RPA approach which neglects the effects of ground-state correlations 
may not be so well founded even for doubly-closed shell nuclei.
There have been attempts to incorporate the effects of ground-state correlations into RPA. The renormalized RPA (rRPA) \cite{rowe1,rowe2} 
includes the fractional occupation of the single-particle states, which plays a role in reducing the particle (p)-hole (h) correlations.
Ground-state correlations not only bring the fractional occupation of the single-particle states
but also give a correlated two-body density matrix. The self-consistent RPA (SCRPA) \cite{scrpa1,scrpa2} includes both the fractional occupation
of the single-particle states and the correlated two-body density matrix which
gives a self energy to a p-h pair and also modifies p-h correlations. 
These rRPA and SCRPA are self-consistent in the sense that the occupation probabilities of the 
single-particle states and the two-body correlation matrix (in the case of SCRPA) are determined using the eigenvectors of the rRPA or SCRPA equations. 
However, the procedures to determine the occupation probabilities and the correlation matrix are complicated and ambiguous \cite{rowe2,scrpa2},
and their applications have been limited to model Hamiltonians. 

The HF+ RPA approach also neglects the coupling to higher amplitudes such as 2p-2h ones.
The coupling to higher amplitudes is essential to describe the damping of collectives excitations such as giant resonances. The second RPA (SRPA) \cite{srpa}
includes
the coupling of the p-h and 2p-2h amplitudes under the assumption of the HF ground state. SRPA has extensively been used to study the damping of collective
excitations \cite{srpa}. It has also been applied to study low-lying dipole states\cite{gamb11,tn}. In the extended SRPA (ESRPA) \cite{srpa}
the effects of ground-state correlations are perturbatively included in the one-body sector of the SRPA equation. ESRPA has been applied
to study the damping of giant resonances in $^{40}$Ca and $^{48}$Ca and it was reported that the effects of ground-state correlations are not significant.
However, the effects of ground-state correlations in other sectors of the SRPA equation and the correlations in the two-body sector are not included in ESRPA. 
Takayanagi {\it et al.} \cite{taka1,taka2} gave a response function formalism which includes the effects of ground-state correlations similarly to ESRPA.

We have developed an extended RPA (ERPA) using the time-dependent density-matrix theory (TDDM) \cite{WC,GT} which consists of the 
coupled equations of motion for one-body and two-body density matrices. The ground state used for ERPA is given as a stationary solution
of the TDDM equations which determine the occupation probabilities of the single-particle states and also the two-body correlation matrix. The ERPA equation
is given as the small amplitude limit of TDDM in a way similar to that RPA is given as the small amplitude limit of the time-dependent HF (TDHF). From the linearization
of the one-body density matrix and the two-body correlation matrix, we obtain a coupled equation for the one-body and two-body transition amplitudes.
Thus the ERPA equation includes all above-mentioned ingredients of extended RPA theories, that is, the fractional occupation of the single-particle states, the two-body correlation matrix 
and the coupling to higher amplitudes.

Using ERPA we have studied the quadrupole excitations of the oxygen isotopes \cite{Toh07} and the dipole and octupole excitations of $^{16}$O
\cite{toh2013}. We found that the two-body correlation matrix which is missing in most extended RPA theories plays a significant role in the collective excitations.
The aim of this paper is to extend our study to heavier doubly-closed shell nuclei $^{40}$Ca and $^{48}$Ca. We calculate the dipole and quadrupole excitations and 
elucidate the effects of the ground-state correlations in these nuclei. Comparing the results for $^{40}$Ca and $^{48}$Ca, we also investigate isotope effects.
For the low-lying states below 10 MeV, we try to compare the existing experimental data \cite{hartmann}. 
The paper is organized as follows. The formulation of ERPA is presented in sect. 2,
numerical details are explained in sect. 3, the obtained results are given in sect. 4 and sect. 5 is devoted to summary.

\section{Formulation}
The TDDM consists of the coupled equations of motion for the one-body density matrix $n_{\alpha\alpha'}$ 
(the occupation matrix) and the correlated part of the two-body density matrix $C_{\alpha\beta\alpha'\beta'}$
(the correlation matrix).
These matrices are defined as
\begin{eqnarray}
n_{\alpha\alpha'}(t)&=&\langle\Phi(t)|a^\dag_{\alpha'} a_\alpha|\Phi(t)\rangle,
\\
C_{\alpha\beta\alpha'\beta'}(t)&=&\langle\Phi(t)|a^\dag_{\alpha'}a^\dag_{\beta'}
 a_{\beta}a_{\alpha}|\Phi(t)\rangle
\nonumber \\
 &-&(n_{\alpha\alpha'}(t)n_{\beta\beta'}(t)
 -n_{\alpha\beta'}(t)n_{\alpha\beta'}(t)),
\end{eqnarray}
where $|\Phi(t)\rangle$ is the time-dependent total wavefunction
$|\Phi(t)\rangle=\exp[-iHt/\hbar] |\Phi(t=0)\rangle$. Here, $H$ is the total Hamiltonian. The equations of motion for reduced density matrices form
a chain of coupled equations known as the Bogoliubov-Born-Green-Kirkwood-Yvon (BBGKY) hierarchy. In TDDM the BBGKY
hierarchy is truncated by replacing a three-body density matrix with anti-symmetrized products of the one-body and
two-body density matrices. The TDDM equation for $C_{\alpha\beta\alpha'\beta'}$ contains all effects of two-body correlations;
p-p, h-h and p-h correlations.
The ground state in TDDM is given as a stationary solution of the TDDM equations. 
The stationary solution can be obtained using the gradient method \cite{Toh07}.
This method is also used in the present work.

The ERPA equations used here are derived as the small amplitude limit of TDDM and are written in matrix form
for the one-body and two-body transition amplitudes $x^\mu_{\alpha\alpha'}$ and $X^\mu_{\alpha\beta\alpha'\beta'}$ \cite{Toh07}
\begin{eqnarray}
\left(
\begin{array}{cc}
a&b\\
c&d
\end{array}
\right)\left(
\begin{array}{c}
x^\mu\\
X^\mu
\end{array}
\right)
=\omega_\mu
\left(
\begin{array}{c}
x^\mu\\
X^\mu
\end{array}
\right),
\label{ERPA}
\end{eqnarray}
where $\omega_\mu$ is the excitation energy.
The effects of ground-state correlations are included in the matrices $a$, $c$ and $d$.
The one-body sector $ax^\mu=\omega_\mu x^\mu$
is formally the same as the RPA equation except for the fact that the occupation matrix $n_{\alpha\alpha'}$ is fractional.
The matrix $d$ also includes $n_{\alpha\alpha'}$, and $c$ both 
$n_{\alpha\alpha'}$ and $C_{\alpha\beta\alpha'\beta'}$. Each matrix element of Eq. (\ref{ERPA}) is given explicitly in Ref. \cite{Toh07}.

Equation(\ref{ERPA}) can be written in the same matrix form as that obtained from
the equation-of-motion approach \cite{rowe1}, which includes the norm matrix
\begin{eqnarray}
\left(
\begin{array}{cc}
S_{1}&T_{1}\\
T_{2}&S_{2}
\end{array}
\right),
\label{norm}
\end{eqnarray}
where $S_1$ is the ground-state expectation value of the commutator between the one-body excitation operators, $T_1~(=T_2^\dag)$ is that between
the one-body and two-body excitation operators and $S_2$ that between the two-body excitation operators.
Using the one-body amplitude $\tilde{x}^\mu_{\alpha\alpha'}$ and
the two-body amplitude 
$\tilde{X}^\mu_{\alpha\beta\alpha'\beta'}$
\begin{eqnarray}
\left(
\begin{array}{c}
x^\mu\\
X^\mu
\end{array}
\right)
=
\left(
\begin{array}{cc}
S_{1}&T_{1}\\
T_{2}&S_{2}
\end{array}
\right)
\left(
\begin{array}{c}
\tilde{x}^\mu\\
\tilde{X}^\mu
\end{array}
\right),
\label{trans}
\end{eqnarray}
we obtain
\begin{eqnarray}
\left(
\begin{array}{cc}
A&B\\
C&D
\end{array}
\right)\left(
\begin{array}{c}
\tilde{x}^\mu\\
\tilde{X}^\mu
\end{array}
\right)
=\omega_\mu
\left(
\begin{array}{cc}
S_{1}&T_{1}\\
T_{2}&S_{2}
\end{array}
\right)
\left(
\begin{array}{c}
\tilde{x}^\mu\\
\tilde{X}^\mu
\end{array}
\right),
\label{ERPA1}
\end{eqnarray}
where $A=aS_1+bT_2$, $B=aT_1+bS_2$, $C=cS_1+dT_2$ and $D=cT_1+dS_2$. 

If the HF assumption is made for the ground state,  Eqs. (\ref{ERPA}) and (\ref{ERPA1}) are reduced to
the SRPA equation \cite{srpa}.

\section{Calculational details}
The occupation probability $n_{\alpha\alpha}$ and the correlation matrix $C_{\alpha\beta\alpha'\beta'}$ are
calculated within TDDM using a truncated single-particle basis depending on the isotope: In the case of $^{40}$Ca
the $2s_{1/2}$, $1d_{3/2}$, $1d_{5/2}$ and $1f_{7/2}$ states are used for both protons and neutrons, while 
the $1f_{5/2}$ states are added in the case of $^{48}$Ca.
For the calculations of the single-particle states we use the Skyrme III force.
To reduce the dimension size, we only consider the 2p-2h and 2h-2p elements of $C_{\alpha\beta\alpha'\beta'}$.
A simplified interaction which contains only the $t_0$ and $t_3$ terms of the Skyrme III force is used as the residual interaction \cite{Toh07}.
The spin-orbit force and Coulomb interaction are also omitted from the residual interaction.
To avoid a rather complicated treatment of the rearrangement effects of a density-dependent force in extended RPA theories \cite{Grasso},
we use the three-body version of the Skyrme interaction,
$v_3=t_3\delta^3({\bm r_1}-{\bm r_2})\delta^3({\bm r_1}-{\bm r_3})$,
which gives the following density-dependent two-body residual interaction \cite{Toh07}:
$t_3\rho_n\delta^3({\bm r}-{\bm r'})$, $t_3\rho\delta^3({\bm r}-{\bm r'})/2$ and 
$t_3\rho_p\delta^3({\bm r}-{\bm r'})$ for the proton-proton, proton-neutron and neutron-neutron interactions, respectively,
where $\rho_p$, $\rho_n$ and $\rho$ are the proton, neutron and total densities, respectively.
The Skyrme III force with the spin-exchange parameter $x_0=0.45$ is strongly attractive in the proton-neutron channel.
Therefore, in the case of $^{48}$Ca where the neutron $1f_{7/2}$ state is fully occupied in the HF approximation, the excitations of the 
2p-2h states which include the neutron $1f_{7/2}$ state become weak and, consequently, the occupation probability of the neutron $1f_{7/2}$ states remains close to
unity and becomes larger than that of the lower-lying neutron $2s_{1/2}$, $1d_{3/2}$, $1d_{5/2}$ states. To avoid such an inversion of the occupation probability, we use
a slightly different interaction for the calculation of the ground state of $^{48}$Ca. 
We neglect the spin exchange parameter $x_0$ and multiply the $t_0$ and $t_3$ parameters with a factor 1.5
so that the correlation energy $E_c$ defined by
\begin{eqnarray}
E_c=\frac{1}{2}\sum_{\alpha\beta\alpha'\beta'}\langle\alpha\beta|v|\alpha'\beta'\rangle C_{\alpha'\beta'\alpha\beta}
\label{Ec}
\end{eqnarray}
becomes similar to the result obtained using the Skyrme III with $x_0=0.45$.

To calculate the excited states, we use Eq. (\ref{ERPA1}) derived from Eq. (\ref{ERPA}) with Eq. (\ref{trans}).
The one-body amplitudes $\tilde{x}^\mu_{\alpha\alpha'}$ are defined using a large number of single-particle states including those in the 
continuum. We discretize the continuum states by confining the wavefunctions in a sphere with radius 15 fm and take all 
the single-particle states with $\epsilon_\alpha\le 50$ MeV and 
$j_\alpha\le 11/2 \hbar$. 
As the residual interaction used in the matrix $a$, $b$ and $c$ in Eq. (\ref{ERPA}), we use a force of the same form as that used in the ground-state
calculation of $^{40}$Ca.
Since the residual interaction is not consistent with the effective interaction used in the calculation of the single-particle states,
it is necessary to reduce the strength of the residual interaction so that the spurious mode corresponding
to the center-of-mass motion comes at zero excitation energy in RPA. We found 
the reduction factor $f$ is 0.66 for $^{40}$Ca and 0.69 for $^{48}$Ca. The residual interaction used in the matrices $a$, $b$ and $c$ is multiplied with this factor $f$.
We include all p-h and h-p amplitudes. For p-p and h-h we include the amplitudes with $|n_{\alpha\alpha}-n_{\alpha'\alpha'}|\ge 0.05$. 
For the quadrupole excitation we also include ${x}^\mu_{\alpha\alpha}$ where $\alpha$ refers to the single-particle states used in the ground-state calculations.
To reduce the number of the two-body amplitudes, we consider only  
the 2p-2h and 2h-2p components of ${X}^\mu_{\alpha\beta\alpha'\beta'}$ using
the 
$2s_{1/2}$, $1d_{3/2}$, $1d_{5/2}$ and $1f_{7/2}$ states for both protons and neutrons in the case of the quadrupole excitation of $^{40}$Ca.
We will show that such small single-particle space is sufficient to obtain a large fragmentation of the quadrupole strength.
The $1p_{1/2}$ and $1p_{3/2}$ states are added for the dipole excitation to create the 2p-2h states with negative parity.
In the case of $^{48}$Ca the $2s_{1/2}$, $1d_{3/2}$, $1d_{5/2}$, $2p_{3/2}$, $1f_{5/2}$ and $1f_{7/2}$
states are used for both protons and neutrons.
Since the single-particle space for
${X}^\mu_{\alpha\beta\alpha'\beta'}$ is similar to that for the calculations of the ground states, 
we use for the matrix $d$ in Eq. (\ref{ERPA}) the residual interactions with the same
form and strength as those used for the ground states of $^{40}$Ca and $^{48}$Ca.


\section{Results}
\subsection{Ground states}
\begin{table}
\caption{Single-particle energies $\epsilon_\alpha$ and occupation probabilities 
$n_{\alpha\alpha}$ calculated in TDDM for $^{40}$Ca.}
\begin{center}
\begin{tabular}{c rr rr} \hline
 &\multicolumn{2}{c}{$\epsilon_\alpha$ [MeV]}&\multicolumn{2}{c}{$n_{\alpha\alpha}$}\\ \hline 
orbit & proton & neutron  & proton & neutron  \\ \hline
$1d_{5/2}$ & -15.6 & -22.9 & 0.923 & 0.924  \\
$1d_{3/2}$ & -9.4 & -16.5 & 0.884 & 0.884  \\
$2s_{2/2}$ & -8.5 & -15.9 & 0.846 & 0.846   \\ 
$1f_{7/2}$ & -3.4 & -10.4 & 0.154 & 0.154  \\ \hline
\end{tabular}
\label{tab1}
\end{center}
\end{table}
\begin{table}
\caption{Same as Table \ref{tab1} but for $^{48}$Ca.}
\begin{center}
\begin{tabular}{c rr rr} \hline
 &\multicolumn{2}{c}{$\epsilon_\alpha$ [MeV]}&\multicolumn{2}{c}{$n_{\alpha\alpha}$}\\ \hline 
orbit & proton & neutron  & proton & neutron  \\ \hline
$1d_{5/2}$ & -22.7 & -22.3 & 0.970 & 0.975  \\
$1d_{3/2}$ & -17.3 & -17.1 & 0.953 & 0.956  \\
$2s_{2/2}$ & -15.1 & -16.4 & 0.902 & 0.960   \\ 
$1f_{7/2}$ & -10.7 & -10.5 & 0.057 & 0.952  \\ 
$1f_{5/2}$ & -2.2 & -1.9 & 0.019 & 0.132  \\ 
\hline
\end{tabular}
\label{tab2}
\end{center}
\end{table}

The occupation probabilities calculated in TDDM for $^{40}$Ca and $^{48}$Ca are shown in Tables \ref{tab1} and \ref{tab2}, respectively.
The deviation from the HF values ($n_{\alpha\alpha}$=1 or 0) is more than 10\% 
in $^{40}$Ca,
which means that the ground state of $^{40}$Ca is a strongly correlated state as was pointed out in the RPA calculation \cite{agassi}.
Because of the occupation of the neutron $1f_{7/2}$ state
the ground-state correlations are weaker in $^{48}$Ca than in $^{40}$Ca.
The correlation energy $E_c$ (Eq. (\ref{Ec})) in $^{40}$Ca is $-21.7$ MeV.
A large portion of the correlation energy is compensated by the increase in the mean-field energy due to the fractional occupation
of the single-particle states. The resulting energy gain due to the ground-state correlations, which is given
by the difference in the total energy between HF and TDDM, is 1.9 MeV. This value is much smaller than $|E_c|=21.7$ MeV. 
In the case of $^{48}$Ca
the correlation energy $E_c$ is $-17.7$ MeV and 
the total energy difference between HF and TDDM is 2.2 MeV. 

\begin{figure} 
\begin{center} 
\includegraphics[height=7cm]{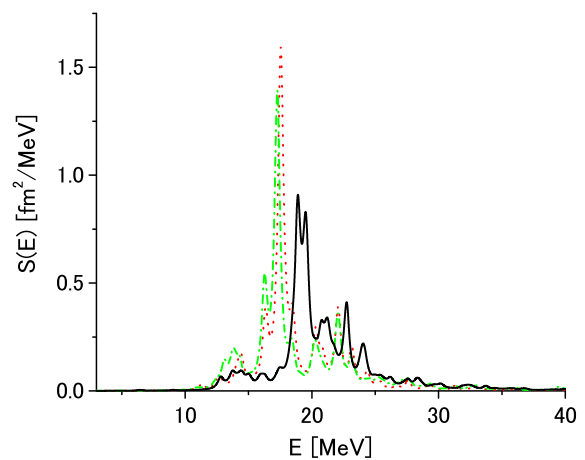}
\end{center}
\caption{Strength functions calculated in RPA (dotted line), SRPA (dot-dashed line) and ERPA (solid line) for the isovector dipole excitation in $^{40}$Ca. 
The distributions are smoothed with an artificial width $\Gamma=0.5$ MeV.} 
\label{e1ca40} 
\end{figure} 
\begin{figure} 
\begin{center} 
\includegraphics[height=7cm]{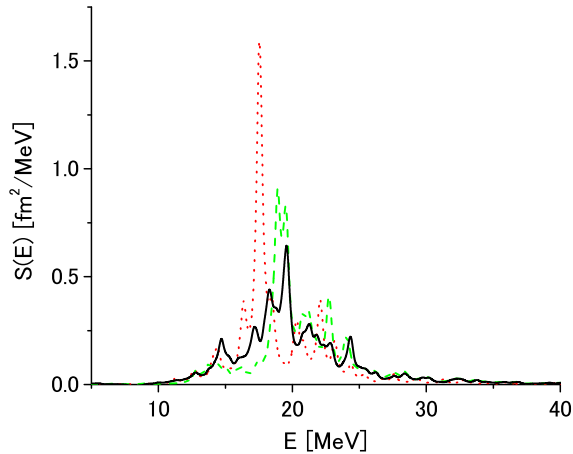}
\end{center}
\caption{Strength function for the isovector dipole excitation in $^{40}$Ca calculated in ERPA (solid line) with the 1p-3h and 1h-3p configurations in addition
to the 2p-2h configurations. The dotted line depicts the result in  RPA and the dot-dashed line the ERPA result including only the 2p-2h configurations. 
The distributions are smoothed with an artificial width $\Gamma=0.5$ MeV.} 
\label{e1ca40m} 
\end{figure} 
\begin{figure} 
\begin{center} 
\includegraphics[height=7cm]{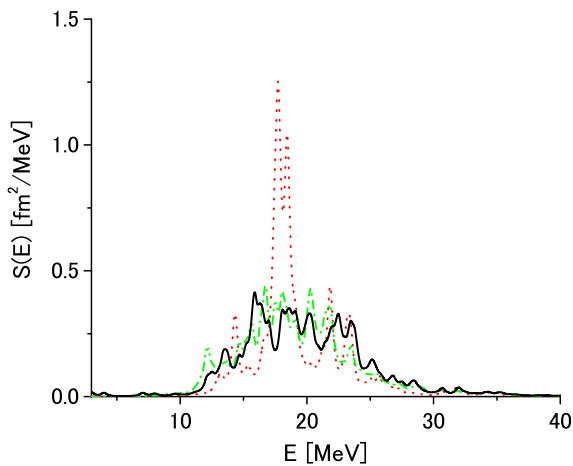}
\end{center}
\caption{Same as Fig. \ref{e1ca40} but for $^{48}$Ca.} 
\label{e1ca48} 
\end{figure}

\subsection{Dipole excitation}
The strength functions for the isovector dipole excitation in $^{40}$Ca calculated in RPA (dotted line), 
SRPA (dot-dashed line), and ERPA (solid line) 
are shown in Fig. \ref{e1ca40}. 
The distributions are smoothed with an artificial width $\Gamma=0.5$ MeV.
The sharp peak in RPA corresponds to the giant dipole resonance (GDR).
In RPA the summed energy-weighted strength exhausts 87\% of the dipole sum rule value including the enhancement term  
which is given by the $t_1$ and $t_2$ parameters of the Skyrme III force.
To increase the the energy-weighted-sum-rule (EWSR) value, we need to improve the residual interaction in a self-consistent manner 
and also the treatment of the continuum states.
The EWSR values of the dipole strength in the other approximations are similar to the RPA value.
From comparison of the results in RPA and SRPA we see that the main effect of the coupling to the 2p-2h configurations
is to shift downward the RPA strength distribution without changing its shape. A similar downward shift of the dipole strength has been reported 
in large scale SRPA calculations \cite{papa,gamb10}.
In ERPA the dipole distribution is upwardly shifted.  
This upward shift is caused by the self-energy contributions \cite{toh2013} in the p-h pairs consisting of the $1d_{5/2}$ and $1f_{7/2}$ states.
The GDR peak is slightly reduced in ERPA but the dipole distribution is not so broad as the experimental data:
The experimental photoabsorption cross section \cite{ahrens,torizuka,diesener} for $^{40}$Ca shows a distribution with width $\Gamma \approx$ 5 MeV.
We performed a slightly larger-scale ERPA calculation using the additional neutron $2p_{3/2}$ and $1f_{5/2}$ states to define the 2p-2h configurations
and found no significant change in the shape of the dipole strength distribution except for a slight further downward shift of the
distribution.
This suggests that the damping of GDR in $^{40}$Ca may be caused by the coupling to other configurations than the 2p-2h ones.
We try to find two-body configurations which have significant values of the matrix elements of $S_2$. When $C_{\alpha\beta\alpha'\beta'}$ is neglected for simplicity, 
the diagonal element of $S_2$ is given by \cite{ts2008}
\begin{eqnarray}
S_2(\alpha\beta\alpha'\beta':\alpha\beta\alpha'\beta')&=&(1-n_\alpha)(1-n_\beta)n_{\alpha'}n_{\beta'}
\nonumber \\
&-&n_\alpha n_\beta (1-n_{\alpha'})(1-n_{\beta'}),
\end{eqnarray}
where we assume that $n_{\alpha\alpha'}=\delta_{\alpha\alpha'}n_\alpha$.
In the case of the HF ground state where $n_\alpha=0$ or 1, $S_2$ is not vanishing only for the 2p-2h configurations:
$S_2$ is 1 (-1) for the 2p-2h (2h-2p) configurations. When the single-particle states are fractionally occupied, all two-body configurations
can have non-vanishing values of $S_2$. Let us assume that $n_{\alpha}=\Delta$ for a particle state and $n_{\alpha}=1-\Delta$
for a hole state
independently of $\alpha$ and that $\Delta$ is small. Then $S_2\approx 1-4\Delta$ for the 2p-2h configurations and 
the 3p-1h (1h-3p) and 3h-1p (1p-3h) configurations have $S_2\approx \Delta (-\Delta)$. $S_2$ for other configurations are
of higher order of $\Delta$. This suggests that the 3p-1h (1h-3p) and 3h-1p (1p-3h) are the next order configurations to be included
in the study of the damping of GDR. The result of such an ERPA calculation is given in Fig. \ref{e1ca40m}, which shows that the spreading of the
dipole strength is enhanced by the coupling to the 3p-1h (1h-3p) and 3h-1p (1p-3h) configurations. This is because these additional configurations have
excitation energies which are close to the GDR energy.

In the case of $^{48}$Ca shown in Fig. \ref{e1ca48}, GDR is strongly damped both in SRPA and in ERPA due to the coupling to the 2p-2h states. 
Since the neutron $1f_{7/2}$ state is occupied in $^{48}$Ca, the 2p-2h states which include the the neutron $1f_{7/2}$ state
as a hole state have energies close to the energy of GDR, resulting in the strong coupling of GDR to the 2p-2h states, 
similarly to the ERPA result for $^{40}$Ca
shown in Fig. \ref{e1ca40m} where the 3p-1h (1h-3p) and 3h-1p (1p-3h) configurations are additionally included.
\begin{figure} 
\begin{center} 
\includegraphics[height=7cm]{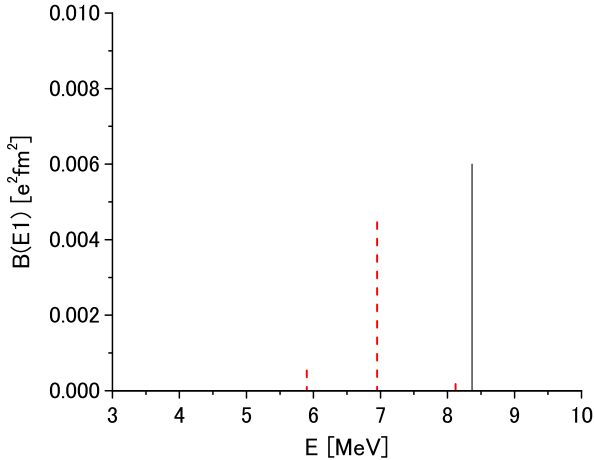}
\end{center}
\caption{Distribution of $B(E1)$ strength below 10 MeV calculated in ERPA for $^{40}$Ca. Experimental data (dashed line) are taken from Ref. \cite{hartmann}.} 
\label{e1ca40low} 
\end{figure} 
\begin{figure} 
\begin{center} 
\includegraphics[height=7cm]{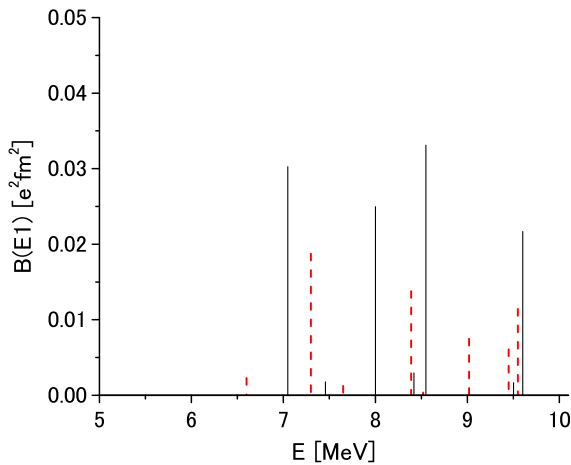}
\end{center}
\caption{Same as Fig. \ref{e1ca40low} but for $^{48}$Ca.} 
\label{e1ca48low} 
\end{figure} 

Since the dipole and quadrupole strength distributions have been measured below 10 MeV using photoluminesence technique \cite{hartmann}, we compare
the dipole strength distributions below 10 MeV calculated in ERPA with experiment \cite{hartmann} 
in Figs. \ref{e1ca40low} and \ref{e1ca48low}.
The inclusion of the additional configurations such as the 3p-1h and 3h-1p configurations increases non-hermiticity of Eq. (\ref{ERPA1}), which makes it
difficult to discuss the dipole states with small transition strength. Therefore, we show in Fig. \ref{e1ca40low} the ERPA result obtained using
the 2p-2h and 2h-2p configurations. The dipole states in ERPA below 10 MeV can have configurations consisting of the p-p transitions 
from the partially occupied particle states
to the continuum states. If we consider only the one-body sector $Ax^\mu=\omega_\mu S_{1}x^\mu$ of Eq. (\ref{ERPA1}), there are no dipole states 
below 10 MeV because the self-energy contributions play an role in upwardly shifting the energies of the p-p pairs \cite{toh2013} 
which have small values of $S_1$ (Eq. (\ref{norm})). 
In Eq. (\ref{ERPA1}) the coupling of the p-p pairs to the 2p-2h or 2h-2p states which
include the same single-particle states as the p-p pairs plays a role in downwardly shifting the p-p pairs. 
Therefore, we include several continuum
states to define the 2p-2h and 2h-2p amplitudes. 
The dipole strength distribution above 10 MeV is little affected by these additional two-body configurations.

In ERPA there is one state at 8.4 MeV in $^{40}$Ca. This dipole state is due to the transition from the partially occupied $1f_{7/2}$ states.
The strength at 8.4 MeV is $7.3$ $\times 10^{-3}e^2$fm$^2$, while the experimental summed strength between
5 and 10 MeV is $5.3\pm0.9$ $10^{-3}e^2$fm$^2$. RPA and SRPA give no dipole states below 10 MeV in $^{40}$Ca. 
A large scale SRPA calculation \cite{papa} for $^{40}$Ca predicts three dipole states below 10 MeV. 
If the residual interaction acting on the matrix $d$ is doubled, our SRPA calculation can also give 
three dipole states below 10 MeV with the summed strength
$2.8$ $\times 10^{-3}e^2$fm$^2$. This is because the 2p-2h states become more fragmented. 
However, such an interaction induces too strong ground-state correlations in TDDM and is not realistic in our scheme.

In the case of $^{48}$Ca there are 7 states below 10 MeV in ERPA.
The two dipole states located between 9 and 10 MeV correspond to the two states in SRPA:
SRPA gives two states at 9.1 and 9.2 MeV for $^{48}$Ca. 
The p-h transition components of the two states exhaust 58 \% 
of the transition amplitudes $(x^\mu,~X^\mu)$.
The other states involve the partially occupied neutron $1f_{5/2}$ state and the p-p transition components exhaust 
$74-83$ \% of each transition amplitude. 
The p-p transitions are not allowed when the ground state is uncorrelated. Therefore, a p-p transition in ERPA may be interpreted 
as a transition from a 2p-2h configuration in the correlated ground state to
a dipole state with another 2p-2h configuration if we assume that the correlated ground state in TDDM consists of the HF state + 2p-2h
configurations.
The summed strength 
below 10 MeV is $103$ $\times 10^{-3}e^2$fm$^2$ while the corresponding experimental value is $61.5\pm7.8$ $\times 10^{-3}e^2$fm$^2$.
The summed strength of the two SRPA states in $^{48}$Ca is only $20\times 10^{-3}e^2$fm$^2$.
Since most of the dipole states below 10 MeV in ERPA are from the single-particle transitions to the continuum states, it is unavoidable
that the strength below 10 MeV somewhat depends on the
size of the sphere used to discretize the continuum states. For example, when a sphere with radius 16 fm is used, the summed strength of 
the dipoles states below 10 MeV in $^{48}$Ca is about 10 \% 
reduced to $90.1$ $\times 10^{-3}e^2$fm$^2$. 
Our RPA calculation gives no dipole states below 10 MeV in $^{48}$Ca. The RPA results below 10 MeV so far reported 
for $^{48}$Ca depend on the effective interaction used.
The density functional theory formalism \cite{chamber} gives some dipole strength below 10 MeV, whereas
an RPA calculation \cite{gamb10} using the SGII force \cite{sgII} does not give dipole states below 10 MeV in $^{48}$Ca.
An RPA calculation for $^{48}$Ca using the same interaction as used in 
the extended theory of finite Fermi systems (ETFFS) \cite{hartmann2} 
gives the dipole strength below 10 MeV which is comparable with the experimental data. 
The large-scale SRPA calculation for $^{48}$Ca performed by Gambacurta {\it et al.} \cite{gamb11} 
predicts a large dipole strength distribution of $230$ $\times 10^{-3}e^2$fm$^2$ below 10 MeV. 
Most of the low-lying dipoles states consists of 2p-2h configurations.
However, such SRPA calculation also gives several MeV downward shift of GDR \cite{gamb11,papa}. 
Therefore, the results of the large-scale SRPA calculations are not conclusive \cite{gamb15}.
ETFFS which includes 1p1h$\otimes$phonon configurations
has also been applied to study the dipole states below 10 MeV in the calcium isotopes \cite{hartmann2}. 
In the case of $^{48}$Ca the number of the low-lying dipole states is increased due to 
the coupling to the 1p1h$\otimes$phonon configurations.

The summed dipole strength below 10 MeV calculated in ERPA for $^{48}$Ca is an order of magnitude larger than that for $^{40}$Ca,
which is in agreement with the experiment \cite{hartmann}. 
Thus ERPA qualitatively describes the difference in the low-lying dipole strength between $^{40}$Ca and $^{48}$Ca,
though it is difficult to predict the excitation energy and transition strength of each dipole state below 10 MeV.
\begin{figure}[h]  
\begin{center} 
\includegraphics[height=7cm]{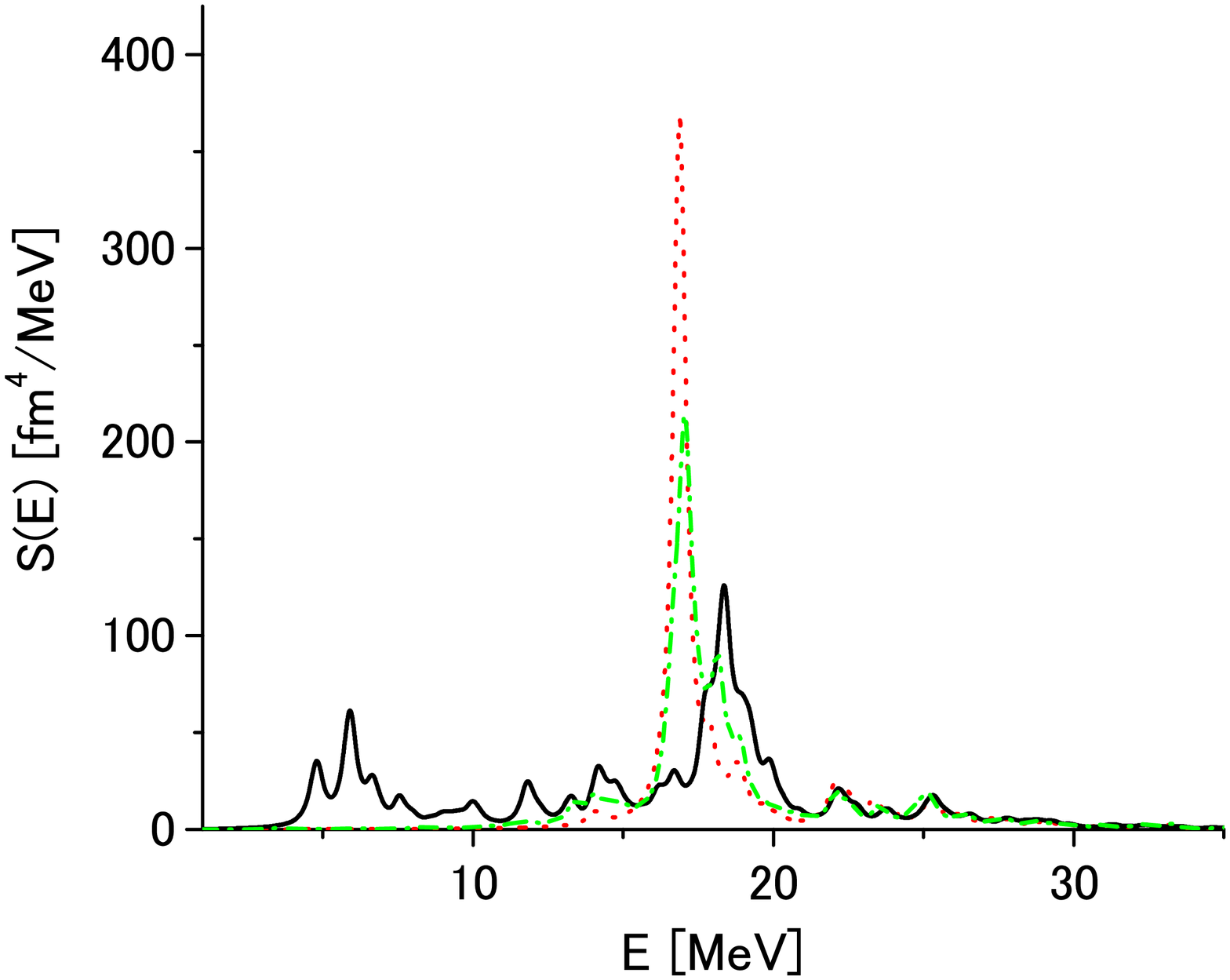}
\end{center}
\caption{Strength functions calculated in RPA (dotted line), SRPA (dot-dashed line) and ERPA (solid line) for the isoscalor quadrupole excitation in $^{40}$Ca. 
The distributions are smoothed with an artificial width $\Gamma=0.5$ MeV.} 
\label{e2ca40} 
\end{figure} 
\begin{figure}[h]
\begin{center} 
\includegraphics[height=7cm]{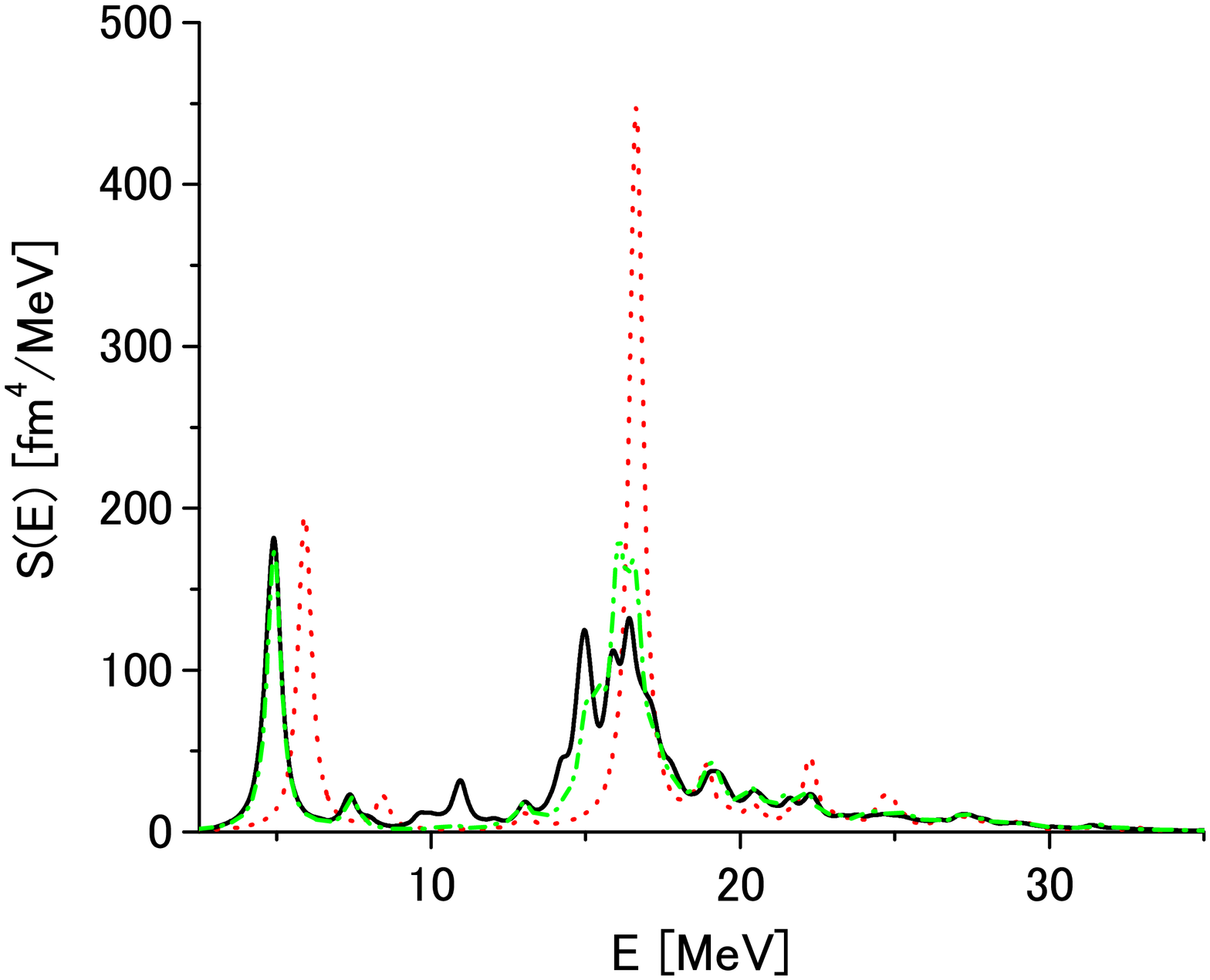}
\end{center}
\caption{Same as Fig. \ref{e2ca40} but for $^{48}$Ca.} 
\label{e2ca48} 
\end{figure}

\begin{figure} 
\begin{center} 
\includegraphics[height=7cm]{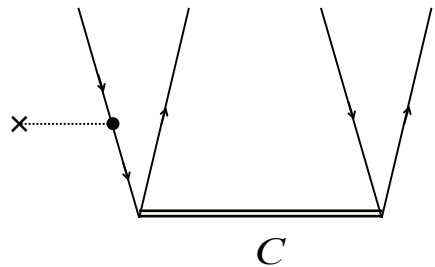}
\end{center}
\caption{Coupling of the h-h amplitude to the 2p-2h amplitude (4 open ended vertical lines) through $C_{\rm pp'hh'}$. 
The horizontal line indicates $C_{\rm pp'hh'}$
and the vertical lines with arrows either a hole state or a particle state. The dotted line with a cross at the left end depicts the
external field and the dot the h-h amplitude.} 
\label{diag} 
\end{figure}

\begin{figure}[h] 
\begin{center} 
\includegraphics[height=7cm]{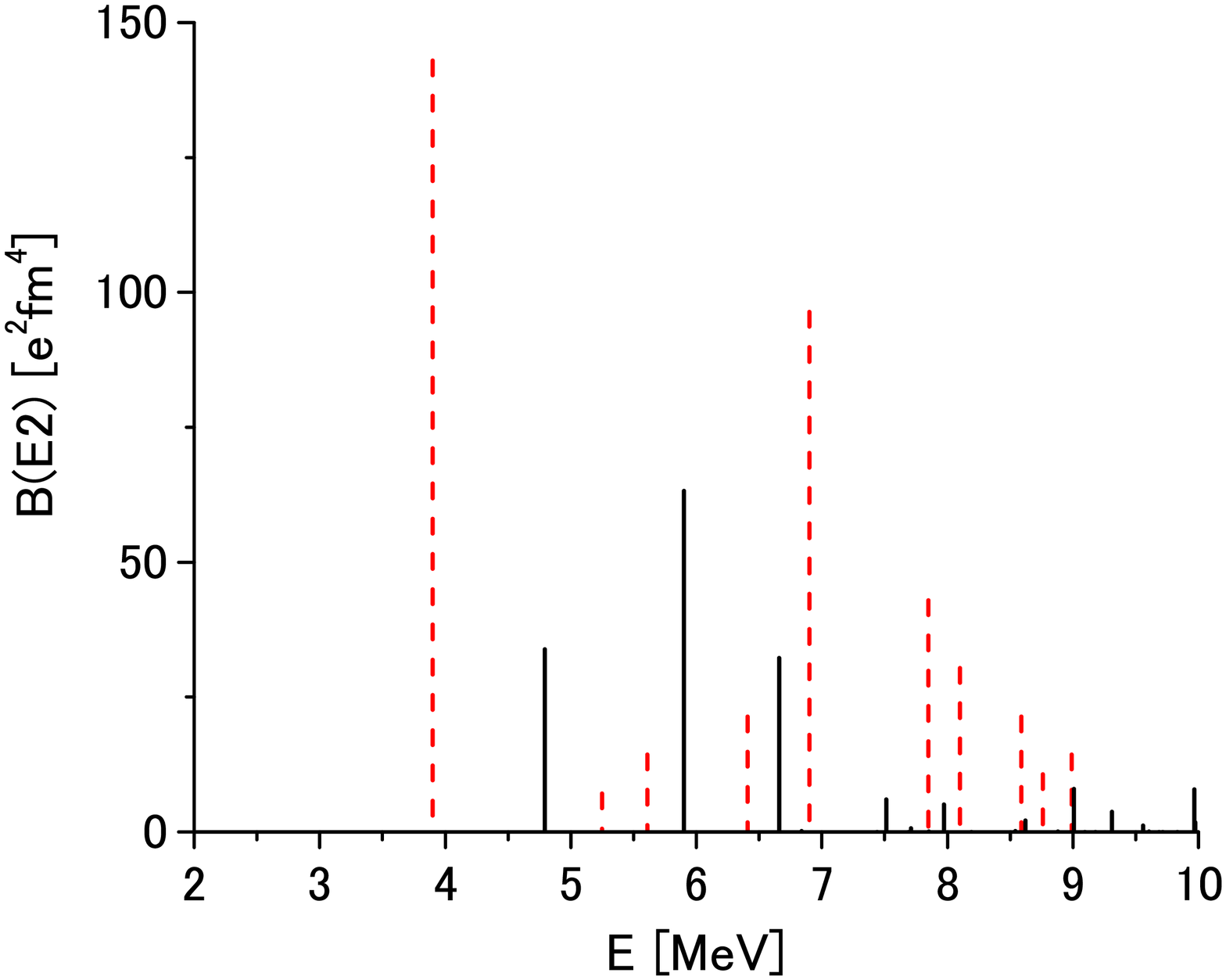}
\end{center}
\caption{Distribution of $B(E2)$ strength below 10 MeV for $^{40}$Ca. Experimental data (dashed line) are taken from Ref. \cite{hartmann}.} 
\label{e2ca40low} 
\end{figure} 
\begin{figure}[h] 
\begin{center} 
\includegraphics[height=7cm]{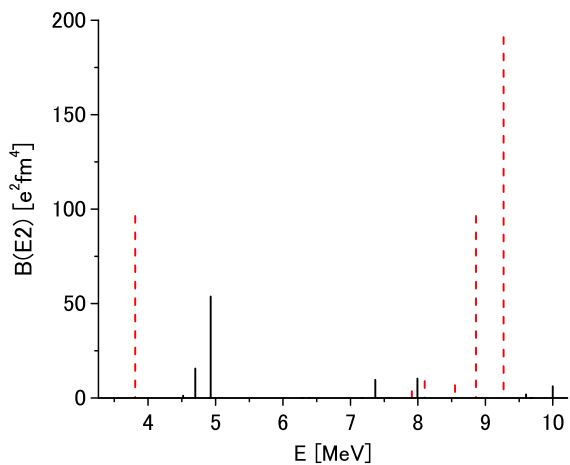}
\end{center}
\caption{Same as Fig. \ref{e2ca40low} but for $^{48}$Ca.} 
\label{e2ca48low} 
\end{figure} 

\subsection{Quadrupole excitation}
The strength functions for the isoscalar quadrupole excitation in $^{40}$Ca and $^{48}$Ca calculated in RPA (dotted line), SRPA (dot-dashed line), and ERPA (solid line) 
are shown in Figs. \ref{e2ca40} and \ref{e2ca48}, respectively. The main peak in RPA corresponds to the giant quadrupole resonance (GQR).
The distributions are smoothed with an artificial width $\Gamma=0.5$ MeV.
In RPA the summed energy-weighted strength exhausts 114\% of the EWSR value. 
To fulfill EWSR, we need a self-consistent treatment of the residual interaction
and the continuum states.
The EWSR values in the other approximations are similar to the RPA value.
From comparison of the SRPA and ERPA results we see that 
the quadrupole strength distribution in $^{40}$Ca is strongly affected by the ground-state correlations.
Especially, a large portion of the quadrupole strength is distributed below 10 MeV and the GQR peak is shifted upward. 
Such large fragmentation of the quadrupole strength was also found in $^{16}$O \cite{Toh07}. The coupling of the 2p-2h amplitudes 
to the h-h or p-p type
amplitudes through  $C_{\alpha\beta\alpha'\beta'}$ 
plays an important role in enhancing the correlations among the 2p-2h configurations \cite{Toh07}. Such a coupling 
is schematically shown in Fig. \ref{diag}. In fact we cannot obtain the strength distribution below 10 MeV
if we neglect the p-p and h-h amplitudes in the ERPA calculations.
The importance of such ground-state correlations on the fragmentation of GQR in $^{40}$Ca was also pointed out by the 1p1h$\otimes$phonon configurations model \cite{kamerd,kamerd2}.
The experimental GQR strength distribution \cite{lisantti} in $^{40}$Ca is split into two peaks. A small peak is located at 14 MeV and a broad one with 
width $\Gamma\approx 5$ MeV at 18 MeV, which is qualitatively reproduced in ERPA.
The quadrupole strength distribution in $^{48}$Ca is moderately affected by the ground-state correlations, 
reflecting the fact that the ground-state correlations are weaker in $^{48}$Ca than in $^{40}$Ca.
The results in SRPA show that the damping of GQR due to the coupling to the 2p-2h configurations is larger in $^{48}$Ca than in $^{40}$Ca.

The strength distributions below 10 MeV calculated in ERPA are compared with experiment \cite{hartmann} in Figs. \ref{e2ca40low} and \ref{e2ca48low}.
In ERPA there are 19 sates below 10 MeV in $^{40}$Ca. 
The first $2^+$ state in $^{40}$Ca mainly consists of 4p-4h states \cite{caurier} as in the case of $^{16}$O \cite{zuker} and cannot be
described by RPA and ERPA.
The summed strength below 10 MeV is $166$ $e^2$fm$^4$ in ERPA, while the corresponding experimental value excluding the
first $2^+$ state
is $263\pm46$ $e^2$fm$^4$. 
In the case of $^{48}$Ca there are 12 states below 10 MeV in ERPA. ERPA cannot reproduce the two states around 9 MeV with strong quadrupole strength.
Consequently the summed strength $28.4$ $e^2$fm$^4$ between 5 and 10 MeV is quite small as compared with the corresponding experimental value
of $302\pm42$ $e^2$fm$^4$. 

\vspace{0.5cm}
\section{Summary}
The effects of ground-state correlations on the dipole and quadrupole excitations were studied for $^{40}$Ca and $^{48}$Ca using the extended
random phase approximation (ERPA) derived from the time-dependent density-matrix theory. 
Large effects of the ground-state correlations were found in the fragmentation of the giant quadrupole resonance in $^{40}$Ca and in the low-lying 
dipole strength below 10 MeV in $^{48}$Ca. It was discussed that the former is due to a mixing of different configurations in the ground state and the 
latter is from the partial occupation of the neutron single-particle states. 
The dipole and quadrupole strength distributions below 10 MeV calculated in ERPA qualitatively agreed with experiment.

\end{document}